\documentclass[twocolumn]{aastex631}
\begin{document}

\newcommand{\Fe}{\ion{Fe}{1}~6302~\AA}
\newcommand{\NaD}{\ion{Na}{1}~D1~5896~\AA}
\newcommand{\ca}{\ion{Ca}{2}~8542~\AA}

\newcommand{\Blos}{\ensuremath{B_\mathrm{LOS}}}

% \title{High-resolution Spectropolarimetric Observation of a C-class Flare by DKIST/ViSP}

%\title{Photospheric Pore Rotation and Flare Triggering from Spectropolarimetric Observations with ViSP/DKIST}

\title{Photospheric Pore Rotation Associated with a C-class Flare from Spectropolarimetric Observations with DKIST}

%\title{Fast Magnetic Pore Rotation During a C-class Flare Observed by DKSIT}

%\title{Spectropolarimetric Analysis of a C-Class Solar Flare from the ViSP Data}
\author{Rahul Yadav}
\affiliation{Laboratory for Atmospheric and Space Physics, University of Colorado, Boulder, CO 80303, USA; \textnormal{rahul.yadav@lasp.colorado.edu}}
%\email{rahul.yadav@lasp.colorado.edu}

\author{Maria D. Kazachenko}
\affiliation{Laboratory for Atmospheric and Space Physics, University of Colorado, Boulder, CO 80303, USA; \textnormal{rahul.yadav@lasp.colorado.edu}}
\affiliation{ National Solar Observatory, 3665 Discovery Drive, 80303, Boulder, CO, USA}
\affiliation{Dept. of Astrophysical and Planetary Sciences, University of Colorado, Boulder, 2000 Colorado Ave, 80305, Boulder, CO, USA}

\author{Andrey N. Afanasyev}
\affiliation{Laboratory for Atmospheric and Space Physics, University of Colorado, Boulder, CO 80303, USA; \textnormal{rahul.yadav@lasp.colorado.edu}}
\affiliation{ National Solar Observatory, 3665 Discovery Drive, 80303, Boulder, CO, USA}
\affiliation{Institute of Solar-Terrestrial Physics of SB RAS, Irkutsk, Russia}
% \altaffiliation{DKIST Ambassador}
\affiliation{DKIST Ambassador}
         
\author{Gianna Cauzzi}
\affiliation{ National Solar Observatory, 3665 Discovery Drive, 80303, Boulder, CO, USA}

\author{Kevin Reardon}
\affiliation{ National Solar Observatory, 3665 Discovery Drive, 80303, Boulder, CO, USA}
\affiliation{Dept. of Astrophysical and Planetary Sciences, University of Colorado, Boulder, 2000 Colorado Ave, 80305, Boulder, CO, USA}

%%%%%%%%%%%%%%%%%%%%%%%%%%%%%%%%%%%%%%%%%%%%%%%%%%%
%%% Abstract 
\begin{abstract}
We present high-resolution observations of a C4.1-class solar flare (SOL2023-05-03T20:53) in AR 13293 from the Visible Spectro-Polarimeter (ViSP)  and Visible Broadband Imager (VBI) instruments at the DKIST. The fast cadence,  good resolution, and high polarimetric sensitivity of ViSP data provide a unique opportunity to explore the photospheric magnetic fields before and during the flare. We infer the magnetic field vector in the photosphere from the \Fe~line using Milne--Eddington inversions. Combined analysis of the inverted data and VBI images reveals the presence of two oppositely polarity pores exhibiting rotational motion both prior to and throughout the flare event. Data-driven simulations further reveal a complex magnetic field topology above the rotating pores, including a null-point-like configuration.
We observed a 30\% relative change in the horizontal component ($\delta F_h$) of Lorentz force at the flare peak time and roughly no change in the radial component. We find that the changes in $\delta F_h$ are the most likely driver of the observed pore rotation. These findings collectively suggest that the back reaction of magnetic field line reconfiguration in the corona may influence the magnetic morphology and rotation of pores in the photosphere on a significantly smaller scale.
 \end{abstract}

%\keywords{ Sun: magnetic fields -- Sun: flares }
\section{Introduction}
Solar flares release energy across the entire electromagnetic spectrum, affecting different layers of the solar atmosphere. They are thought to be driven by the rapid release of magnetic energy stored in the magnetic fields of active regions (ARs) in the corona.
Several models have been proposed to understand solar flare evolution, but the mechanism responsible for flare trigger remains unclear \citep{2011SSRv..159...19F, 2022SoPh..297...59K}. According to the CSHKP model \citep{1964NASSP..50..451C, 1966Natur.211..695S, 1974SoPh...34..323H, 1976SoPh...50...85K}, one possibility is that solar flares are caused by the magnetic reconnection in the corona.

Although a solar flare can occur in almost any AR, observations have demonstrated that ARs with a complex magnetic structure in the photosphere are more likely to produce flares \citep{2009AdSpR..43..739S, 2019LRSP...16....3T}. 
Several studies have also reported that the emergence of magnetic flux or photospheric shearing motions can trigger flares, either via interaction of new flux with preexisting field lines or by twisting the coronal field lines via motions of their footpoints anchored in the photosphere \citep{2001ApJ...552..833M,2008A&A...492L..35A, 2015SoPh..290.3641L, 2019ApJ...871...67C, 2023ApJ...955..105R,2023ApJ...955...34Q,2024ApJ...962..149T}.

Observations, particularly during intense X- or M-class flares, have also shown that flares can directly impact the photosphere, resulting in rapid sunspot rotation \citep{1993SoPh..147..287A,2014ApJ...782L..31W, 2016NatCo...713798B} and a significant increase in horizontal magnetic fields near the polarity inversion line(PIL; \citealt{2005ApJ...635..647S,2017ApJ...839...67S, 2019ApJS..240...11P, 2021A&A...649A.106Y, 2023ApJ...944..215Y}). For instance, \cite{2016NatCo...713104L} reported sudden flare-induced non-uniform rotation of a sunspot ($\approx50^\circ$h$^{-1}$) using high-resolution observations. Generally, the sunspot rotation is associated with the back reaction of the flare-related restructuring of the coronal magnetic field \citep{2008ASPC..383..221H, 2012SoPh..277...59F,2016NatCo...713104L}. 

In this study, we focus on the rotation of two photospheric pores associated with a C4.1-class solar flare observed by the 4 m Daniel K. Inouye Solar Telescope \citep{2020SoPh..295..172R}. Utilizing the high polarimetric sensitivity and spectral resolution of the Visible Spectro-Polarimeter (ViSP; \citealt{2022SoPh..297...22D}), we aim to understand the cause of the observed rotation  and its connection to the C4.1-class solar flare.

In Section \ref{Sec:observations} we describe our observations and data reduction. Sections \ref{sec:method_and_data}
 and \ref{sec:results} present our analysis and results. The obtained results are discussed in Section \ref{sec:Discussion}, and summarized in Section~\ref{sec:conclusion}.
%%%%%%%%%%%%
\section{Data Acquisition, Reduction, and Overview}
\label{Sec:observations}
\subsection{Data Acquisition}
\begin{figure*}[!ht]
    \centering
    \includegraphics[clip,trim=3.4cm 0.4cm 3.5cm 0.1cm,width=1.\textwidth]{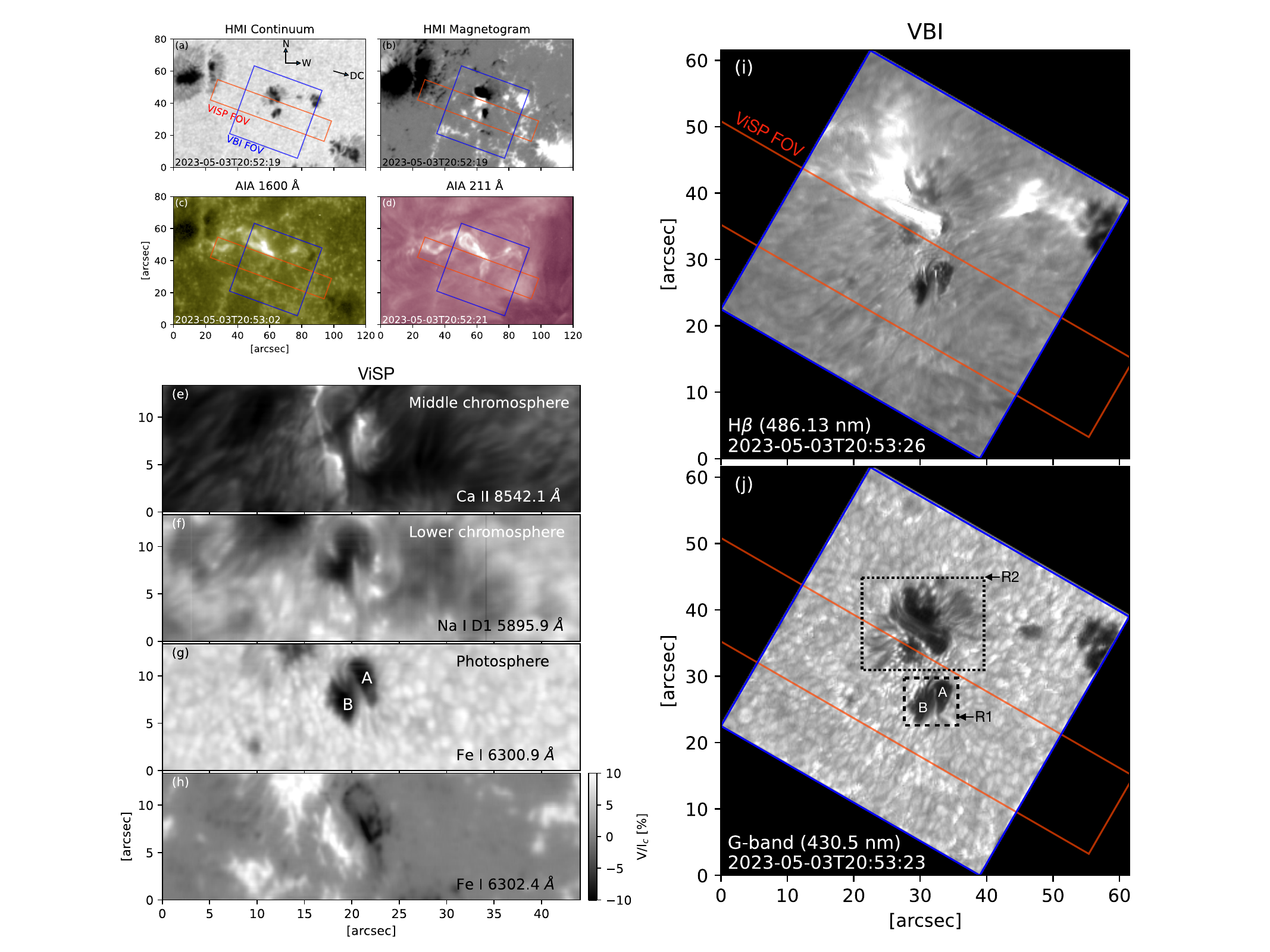}
    \caption{Overview of the C4.1-class flare (SOL2023-05-03T20:53) in AR 13293. Panels (a)--(d): HMI continuum image and LOS magnetogram, AIA 1600 and 211~\AA$ $ images observed by the SDO. Red and blue rectangles outline the FOV covered by ViSP (arm1) and VBI, respectively. Panels (e)--(h): coaligned ViSP FOV in three arms at 20:09 UT on May 3, 2023 centered on the two magnetic pores (A and B). (e) Line core intensity map in arm \ion{Ca}{2}~8542~\AA, (f) line core intensity map in arm \ion{Na}{1}~D1~5896~\AA, (g) continuum intensity map, and (h) Stokes V signal in the photosphere at wavelength \ion{Fe}{1}~6302.4~\AA. Panels (i)--(j) VBI H$\beta$ and G-band intensity maps. R1 and R2 highlight regions with counterclockwise and clockwise rotation, respectively. An animation of the VBI images (panels (i)--(j)) is available in the online Journal. The VBI animation, which does
not include the annotations shown in the static figure, proceeds from 18:43 to 21:13 UT on 2023-05-03..}
    \label{fig:hmivisp_fov}
\end{figure*}

In Figure~\ref{fig:hmivisp_fov}, we show an overview of the C4.1-class flare (SOL2023-05-03T20:53) that was captured by ViSP and the Visible Broadband Imager
(VBI; \citealt{2021SoPh..296..145W}) installed at DKIST. This flare occurred in AR NOAA 13293 at 20:47 UT, peaking at 20:53 UT and ending at 20:58 UT on May 3, 2023. ViSP and VBI observed this AR from 18:43 to 21:27 UT, capturing the entire flare duration. Due to the trade-off between the number of scan steps and the scan repetition cadence, we recorded only a portion
of the AR with the ViSP. The center of the ViSP field of view (FOV) was located at helioprojective coordinates X$=-$580\arcsec~and Y$=255$\arcsec~(N13E37). 

ViSP recorded three spectral regions simultaneously with three separate detectors. The observing program of ViSP was designed to record repeated scans of the FOV across the \ion{Fe}{1} line at 6302~\AA\ in arm~1, the \ion{Na}{1} D2 line at 5896~\AA\ in arm~2, and the \ion{Ca}{2} line at 8542~\AA\ in arm~3. In our data, the slit size was 0.107\arcsec.
The spatial scan was performed along the perpendicular direction of the slit in a total of $125$ steps, resulting in an FOV of 1$3.4\arcsec \times$  slit length, where the slit lengths for 6302~\AA, 5896~\AA, and 8542~\AA\ channels are 75.5\arcsec, 61\arcsec, and 49\arcsec, respectively. The spectral sampling  of arm 6302~\AA, arm 5896~\AA, and arm  8542~\AA\ are 12.8 m\AA, 14~m\AA, and 18.8 m\AA, respectively. At each slit step position, the data were acquired in full polarimetry mode with four modulation cycles and ten modulation states. The total integration time per slit step was $1.5$ s, resulting in sequences of $48$ raster maps with a cadence of $3.11$ minutes.

For calibration purposes, we used the same observing program to record three raster maps of the quiet-Sun region between 21:41 and 21:51 UT on May 3, 2023 at the disk center. 

In addition to ViSP, the VBI observed quasi-simultaneously high-resolution (0.01\arcsec~per pixel) intensity
images in the \textit{G} band and H$\beta$ with a cadence of 9~s. Level-1 ViSP data, as well as speckle-reconstructed VBI data, were provided by the DKIST Data Center. The DKIST Data Center Archive hosts these observations under proposal identifier as pid\_2\_64.
\subsection{Data Reduction}
ViSP Level-1 data include initial corrections for instrumental effects and polarimetric calibration \footnote{\url{https://docs.dkist.nso.edu/projects/visp/en/v2.0.1/index.html}} to perform an initial correction for instrumental effects and polarimetric calibration. For scientific analysis, we had to further process the data to account for wavelength calibration, removal of residual crosstalk, and alignment of all maps. Absolute wavelength calibration of spectral lines was conducted by comparing the mean spectra of the quiet Sun observed at the solar disk center with the BASS-2000 solar atlas profiles\footnote{\url{https://bass2000.obspm.fr/solar_spect.php}}. 

To remove the crosstalk in the \ion{Fe}{1} spectra, we combined the approaches provided by \cite{1992ApJ...398..359S} and \cite{1994SoPh..153..143K}. We first determined the crosstalk terms from Stokes $I$ to $Q$, $U$, and $V$ based on the technique of \cite{1992ApJ...398..359S}. We then estimated the crosstalk terms from Stokes $V$ to $Q$ and $U$ using the approach provided by \cite{1994SoPh..153..143K}. 
By employing this approach, we believe that residual crosstalk is below the noise level of the data. To improve the signal-to-noise ratio and to get a square pixel of 0.107\arcsec, we rebinned the ViSP pixels along the slit, resulting in a noise level 
of $\sim8 \times 10^{-4}$ relative to the continuum in the photosphere.  
We then used the continuum intensity images and the  the cross-correlation approach to coalign the images in time.

\subsection{Overview of Observations}
In Figure~\ref{fig:hmivisp_fov} we show an overview of the analyzed C4.1-class flare from different instruments; panels (a)--(d) show contextual images from the  Helioseismic and Magnetic Imager (HMI; \citealt{2012SoPh..275..207S}) and the Atmospheric Imaging Assembly (AIA; \citealt{2012SoPh..275...17L}) on board of the Solar Dynamic Observatory (SDO; \citealt{2012SoPh..275....3P}). 
A red rectangle  shows the ViSP FOV that covers two small pores of opposite polarities and a small part of the penumbra located above the pores, whereas a blue box outlines the FOV of VBI.
The VBI FOV covers the ribbon locations as well as the two pores (A and B) covered by ViSP (see panels i and j of Figure~\ref{fig:hmivisp_fov}).

The HMI magnetogram shows a complex structure of the AR, which can be seen more clearly in the VBI images.
The R1 and R2 regions, highlighting two spots of mixed polarity in close proximity, rotate clockwise and counterclockwise during our DKIST observations, respectively. Prior to the main C4.1-class flare, these regions exhibited small-scale brightenings in H$\beta$ images (see available animation). The footpoint of the C4.1 flare is located in the R2 region, near highly sheared penumbral fibrils separating the polarities. Additionally, a smaller flare (C1.9 class) occurred at 18:49 UT, associated with the R1 region (pores A and B).

Figure \ref{fig:hmivisp_fov}(e-h) shows intensity maps of the ViSP FOV in three arms covering a range of altitudes from the photosphere to middle chromosphere, as well as the map of Stokes V (circular polarization) for the photospheric line. The continuum intensity in the \ion{Fe}{1} 6302 line pair (Figure \ref{fig:hmivisp_fov}g and h) shows granules , two opposite polarity pores, and part of the sunspot structure. The line core intensity map of \ion{Na}{1} D1 line (Figure \ref{fig:hmivisp_fov}f) shows typical features in the lower photosphere/upper chromosphere \citep{2000A&A...357.1093C,2010ApJ...709.1362L, 2016ApJ...832..147K}. Strong intensity enhancement of flare ribbons can be seen in the line core of \ion{Ca}{2} 8542 Å (Figure \ref{fig:hmivisp_fov}e), which forms in the middle chromosphere \citep{2008A&A...480..515C, 2017SSRv..210..109D}.
The simultaneous observations of these three lines, forming in different layers of the solar atmosphere, can provide information on how the atmosphere behaves throughout the lower solar atmosphere during flares. 

However, in the rest of this Letter we focus on the \Fe~line only, leaving multiline photospheric and chromospheric analysis for future work. 

\begin{figure*}[!ht]
    \centering
    \includegraphics[width=1\textwidth]{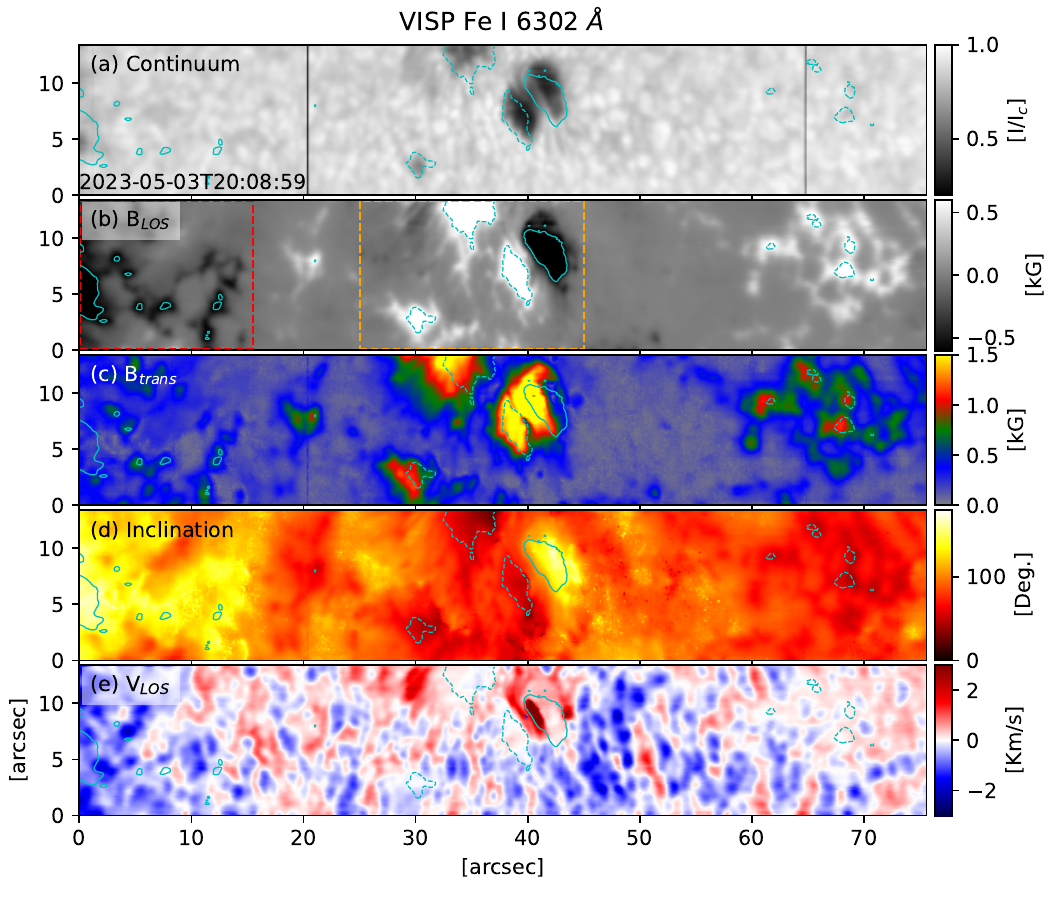}
    \caption{Photospheric magnetic field properties and LOS velocity derived from the Milne--Eddington inversion of \ion{Fe}{1}~6302~\AA\ ViSP/DKIST line. (a) Continuum intensity map, (b) LOS magnetic field, (c) transverse component of magnetic field, (d) inclination angle, and (e) LOS velocity in the photosphere. Overlaid solid and dashed contours refer to the LOS magnetic field strength of $\pm0.6$~kG, respectively. The red and orange boxes highlight the FOVs used for the HMI and ViSP comparison. An
animation of Fe I 6302~\AA~ViSP/DKIST line (all panels) is available in the online Journal. The ViSP animation,
which does not include FOVs shown in the static figure, proceeds from 18:43 to 21:07 UT on 2023-05-03.}
    \label{fig:ME_fena}
\end{figure*}

\begin{figure*}[!ht]
    \centering
    \includegraphics[width=0.5\textwidth]{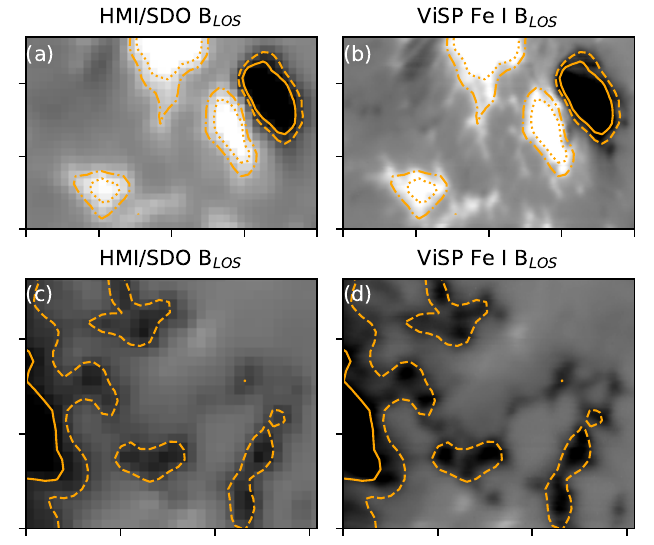}
    \includegraphics[width=0.45\textwidth]{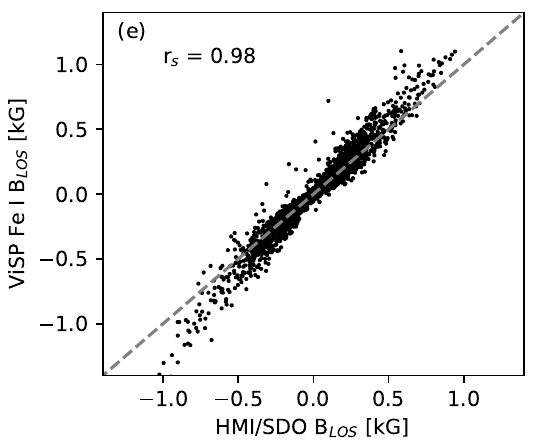}
    
    \caption{Comparison of HMI ((a),(c)) and ViSP ((b),(d)) LOS magnetic field highlighted within the orange and red boxes in Figure~\ref{fig:ME_fena}. HMI LOS magnetic field contours are overlaid on the ViSP and HMI FOVs, the difference between two consecutive ticks corresponds to 5\arcsec. Solid, dashed, dotted, and dotted-dashed contours refer to the values of $-500$~G, $-300$~G, 500~G, and 300~G, respectively. (e) Scatterplot of LOS magnetic field derived from ViSP vs HMI. r$_s$ is the Spearman correlation coefficient.} 
    \label{fig:visp_hmi_blos_compare}
\end{figure*}
\section{Methods and Data 
Analysis}\label{sec:method_and_data}
\subsection{Inversion of the Spectropolarimetric Data} 
We invert individual ViSP Fe~I lines (i.e., \ion{Fe}{1}~6302.5~\AA) using the SPIN Milne--Eddington inversion code \citep{2017SoPh..292..105Y} to retrieve the magnetic field vector and line-of-sight (LOS) velocity in the photosphere. 
We then use the automated ambiguity resolution code of \citet{2014ascl.soft04007L} to resolve the 180$^{\circ}$ azimuthal ambiguity. This method relies on the minimum energy method by \citet{1994SoPh..155..235M}. Finally, we use the transformation matrix by \cite{1990SoPh..126...21G}  to convert the magnetic field vector inferred in the LOS frame to the solar local reference frame.

\subsection{Lorentz Force on the Solar Atmosphere}
We estimate the changes in the Lorentz force in the photosphere using the method by \cite{2012SoPh..277...59F}. They showed that the Lorentz force acting on the photosphere can be approximated by a surface integral using photospheric boundary data alone. The temporal changes in the radial and horizontal components of the Lorentz force are then
\begin{equation}
    \delta F_r = \frac{1}{8\pi}\int{dA (\delta B_r^2- \delta B_h^2}),
    \label{Equation1}
\end{equation}
\begin{equation}
\delta F_h = \frac{1}{4\pi}\int{dA\delta(B_rB_h)},
\label{Equation2}
\end{equation}
where $\delta F_r$ and $\delta F_h$ are the change in radial and horizontal
components of the Lorentz force, respectively, over the surface area $dA$ in the photosphere. $B_h$ and $B_r$ are the horizontal and radial components of magnetic field in the photosphere. $\delta B_h$ and $\delta B_h$ are the temporal changes in the horizontal and radial component of magnetic field.

\subsection{Data-driven Simulation of the AR 13293}
\label{sec:cgem_intro}
To derive the 3D magnetic field evolution in AR 13293, we ran the Coronal Global Evolutionary Model (CGEM; \citealt{2020ApJS..250...28H}) from 2023 May 2, 07:54 UT, to May 3, 21:42 UT. As an input, CGEM utilizes electric field maps derived from photospheric vector magnetograms and Dopplergrams \citep{2020ApJS..248....2F} to create a time-dependent, magnetofrictional (MF) model of nonpotential magnetic field \citep{2012ApJ...757..147C,Lumme2022}. 
We use the HMI data instead of ViSP to perform the CGEM/MF simulations because the ViSP FOV is limited and does not satisfy the flux-balance condition required for our simulation.
Data from the HMI/SDO provide the photospheric magnetic fields and Doppler velocity, while the ``PDFI'' inversion method \citep{2014ApJ...795...17K, 2020ApJS..248....2F} computes the electric field. Our 3D domain has 449$\times$417$\times$321 grid points at a 0\arcsec.5 resolution.  

Our focus in this simulation was to understand the magnetic field topology within the AR using data-driven simulations. We do not aim to analyze flare signatures or compare them with observations.

%%%%%%%%%%%%%%%%%%%
\section{Results}
\label{sec:results}
\subsection{Comparison of Magnetic Fields from  ViSP/DKIST and HMI/SDO}
Figure \ref{fig:ME_fena} shows maps of the longitudinal (\Blos) and transverse ($B_{trans}$) components of the magnetic field, inclination angle, and LOS velocity in the photosphere that were derived from the Milne--Eddington inversions of ViSP \ion{Fe}{1}~6302~\AA~spectral line observations. We find that the average value of the LOS magnetic field in the two pores, A and B as labeled in Figure~\ref{fig:hmivisp_fov}g, is $\sim$1.2~kG in the photosphere. 
The transverse component of the magnetic field also exhibits strong values ($\sim$1.5~kG) in the pore and sunspot regions. The LOS velocity distribution ranges within $\pm$2~km~s$^{-1}$, which is typical of photospheric observations. We also observe a strong downflow close to the PIL between the pores with a mean value of 3.9~km~s$^{-1}$. 

For comparison purpose, we used the HMI \Blos~data with a 12 minute cadence. We note that the HMI and ViSP data have different cadences, with the ViSP having a higher cadence of $3.11$~minutes. HMI data obtained at 20:12 UT on 2023 May 3 were compared with ViSP data observed around the same time.

In Figure \ref{fig:visp_hmi_blos_compare}, we compare the \Blos~obtained from the ViSP \Fe~line with the \Blos~obtained from the HMI instrument on board SDO using the \ion{Fe}{1}~6173~\AA\ spectral line. The HMI utilizes an imaging-based spectropolarimeter for polarimetric observations, while ViSP is a slit-based spectropolarimeter. Despite the differences in the polarimetry techniques and spectral lines used by the two instruments (\ion{Fe}{1}~6173~\AA~and \Fe), they both observe at roughly the same height in the photosphere \citep{2017ApJ...851..111S}. 

We find a remarkable similarity between the \Blos~obtained from two different instruments. Figure~\ref{fig:visp_hmi_blos_compare} demonstrates that the ViSP data better resolve small-scale magnetic structures compared to HMI, where same features appear more pixelated.

Furthermore, we do pixel-by-pixel comparison between \Blos~obtained from the ViSP~\ion{Fe}{1} and HMI (see Figure~\ref{fig:visp_hmi_blos_compare}e). Since the spatial resolution of our ViSP data is approximately 5 times better than HMI's, we resample the ViSP FOV down to HMI's resolution before \Blos~comparison. We find a strong correlation ($r_s = 0.98$) between the \Blos~values across the entire ViSP FOV. A similar correlation coefficient ($r_s \sim 0.95$) value
was also found in the ViSP and HMI comparison by \cite{2023ApJ...954L..35D}, but for a plage region. In our data, the difference between ViSP and HMI increases in stronger field regions, with a mean difference of $\approx$~200~G.  
This is likely due to the differences in the cadence, spatial and spectral resolutions of the two instruments. Additionally, our ViSP data at a spatial resolution of 0.1\arcsec~reveals that few pixels in the negative pore harbor strong \Blos~values ($\sim$~2~kG), which are not observed in the HMI magnetogram. 

\begin{figure*}[!t]
    \centering
    \includegraphics[clip,trim=6cm 2.6cm 9cm 0cm,width=1\textwidth]{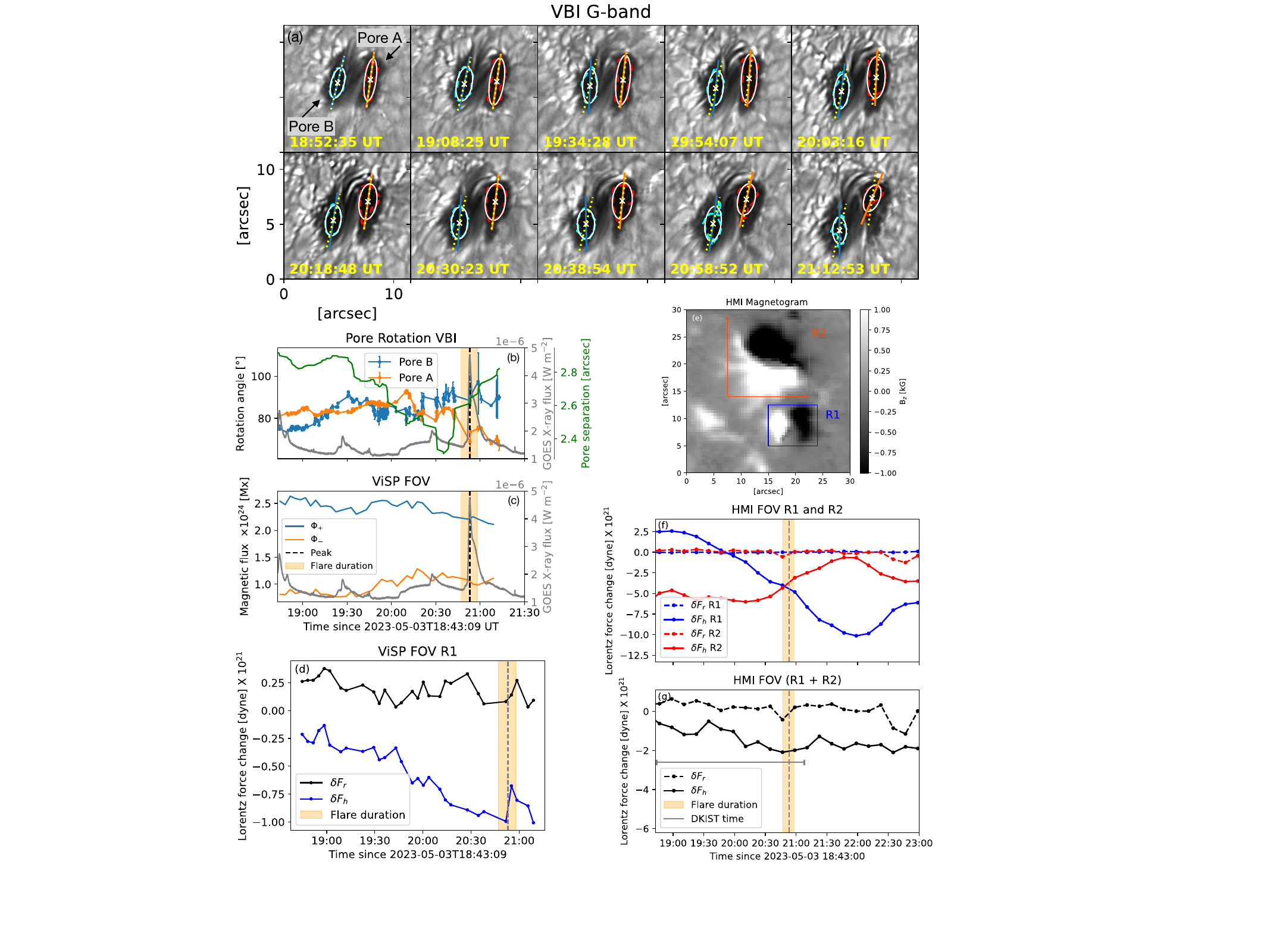}
    \caption{Pores' rotation derived from VBI images and the changes in Lorentz force within the ViSP and HMI FOVs.
    Panel (a): pore rotation in VBI \textit{G} band (red: pore A, cyan: pore B). White ellipses represent the best fit for each pore's boundary; yellow dashed line indicates a reference angle. Panel (b): temporal evolution of pores' rotation and the separation between them derived from VBI images. Panel (c): temporal evolution of positive ($\phi_{+}$) and negative ($\phi_{-}$) magnetic fluxes in the ViSP FOV. The GOES X-ray flux (1--8~\AA) is shown in the gray line. Panel (d): temporal evolution of horizontal ($\delta F_h$) and radial ($\delta F_r$) component of Lorentz forces in the ViSP FOV (consisting of Pore A and B; R1 region highlighted in panel~e). Panels f and g: Temporal evolution of $\delta F_h$ and $\delta F_r$ derived from HMI in the regions R1 and R2 highlighted in panel e.}
    \label{fig:pore_rot_flux}
\end{figure*}

\subsection{Rotation of Pores in the Photosphere from VBI}
From the ViSP and G-band data, we find that two pores of the opposite polarity, pore A and B in Figure~\ref{fig:hmivisp_fov}g and j, rotate counterclockwise in the period leading up to, and during, the flare. The VBI G-band images have better spatial resolution and temporal cadence compared to the ViSP continuum images. Therefore, to estimate pore rotation, we apply an intensity threshold to quiet-Sun normalized VBI G-band images to define the area for elliptical fitting. In this estimate, we omit the images affected by worse seeing conditions. 
From the ellipse fitting, we derived the orientation, centroid, semi-major, and semi-minor axes of the ellipse. We define the zero angle of the ellipse to be aligned with the x-axis of Figure~\ref{fig:hmivisp_fov}j.

In addition to pore rotation, VBI G-band images demonstrated that the mixed polarity sunspot located above pores A and B exhibited clockwise rotation (see Figure~\ref{fig:hmivisp_fov}j and animation). Due to its complex shape, the ellipse fitting approach was not applicable to that region. In this Letter, we focus on the rotation of pores, which are also covered by ViSP.

Figure \ref{fig:pore_rot_flux}(a)~and~(b) show the rotation angle derived from ellipse fitting for each pore before and during the flare. We find that pore B exhibited rapid rotation of 17$^{\circ}$ from 19:00 to 19:30 UT with an angular speed of 34$^{\circ}$~hr$^{-1}$, preceding the C4.1-class flare. During the flare, the pore rotated an additional 10$^{\circ}$ from 20:30 to 21:00 UT at a rate of 20$^{\circ}$~hr$^{-1}$. In contrast, pore A displayed a more gradual rotational increase, rotating at a slower pace of approximately 9$^{\circ}$~hr$^{-1}$ from 18:45 to 20:10 UT.

After 20:45 UT, when the C4.1-class flare occurred, the reduced number of usable frames due to degraded seeing conditions led to more fluctuations and uncertainties in the estimated rotation angles. Another possibility for the uncertainties in rotation angles could be the worse ellipse fitting of the pore boundary due to its nonellipse shape (see available animation). To estimate the error in pore rotation, we performed ellipse fitting using a different intensity threshold, which was $5\%$ relative (lower) to the quiet-Sun normalized reference intensity threshold value ($\sim$0.34).

We also find that the separation distance between pores A and B, calculated using their centroids, reduced by 0.6\arcsec~between 19:00 and 20:35 UT, resulting in a convergence speed of approximately 68~m~s$^{-1}$. Before flare onset (after 20:35 UT), the two pores began separating at a speed of approximately 146~m~s$^{-1}$ (green curve in Figure~\ref{fig:pore_rot_flux}). This suggests that, in addition to rotation, this region also exhibits shearing motion in the photosphere. 

The temporal evolution of magnetic flux estimated from the ViSP FOV ( Figure~\ref{fig:pore_rot_flux}(c)), indicates that there is no significant emergence of new magnetic flux within the ViSP FOV during our observation period.

\subsection{Pore Rotation and Lorentz Force Change}
\begin{figure*}
    \centering
    \includegraphics[width=0.30\textwidth]{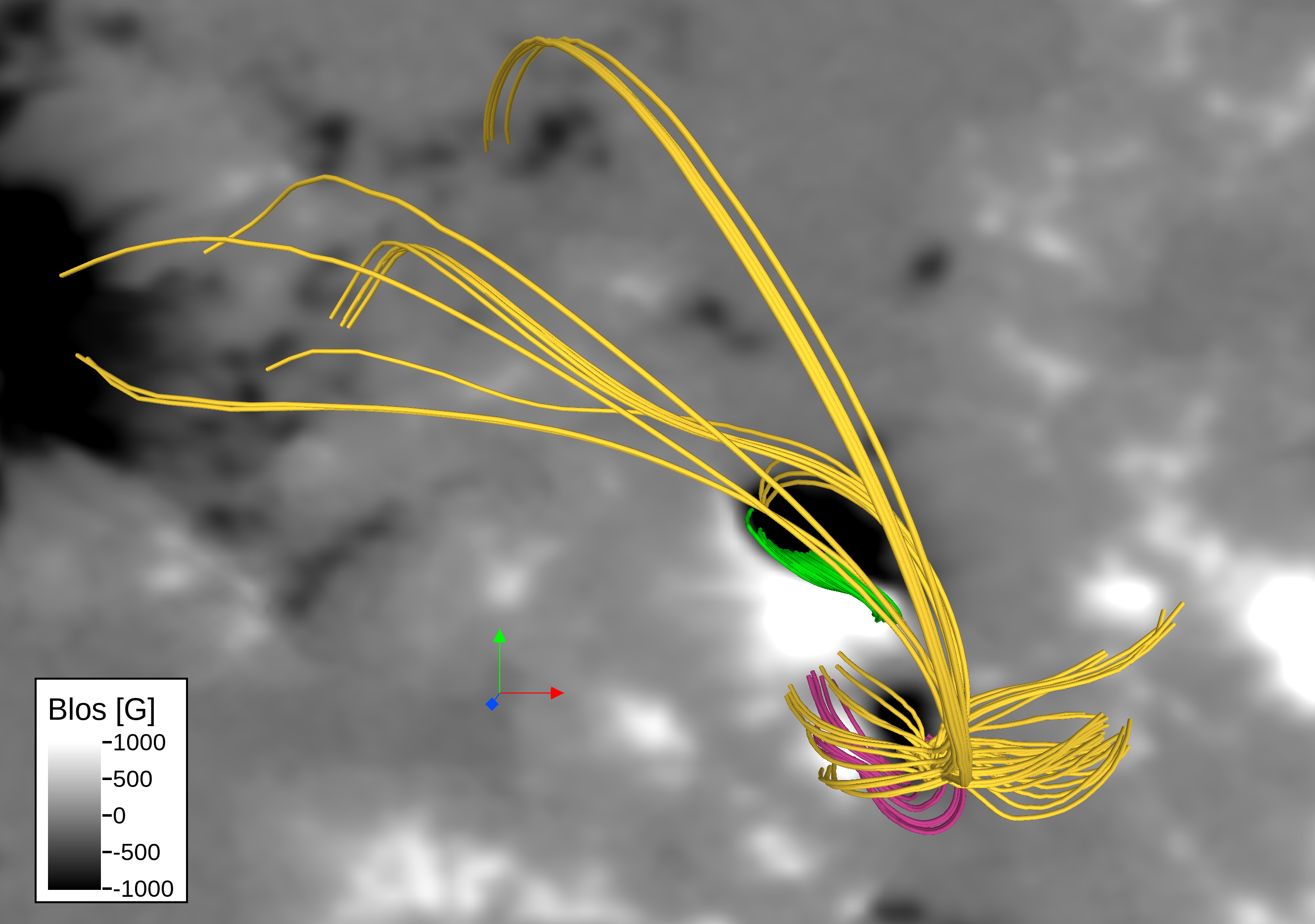}
    \includegraphics[width=0.33\textwidth]{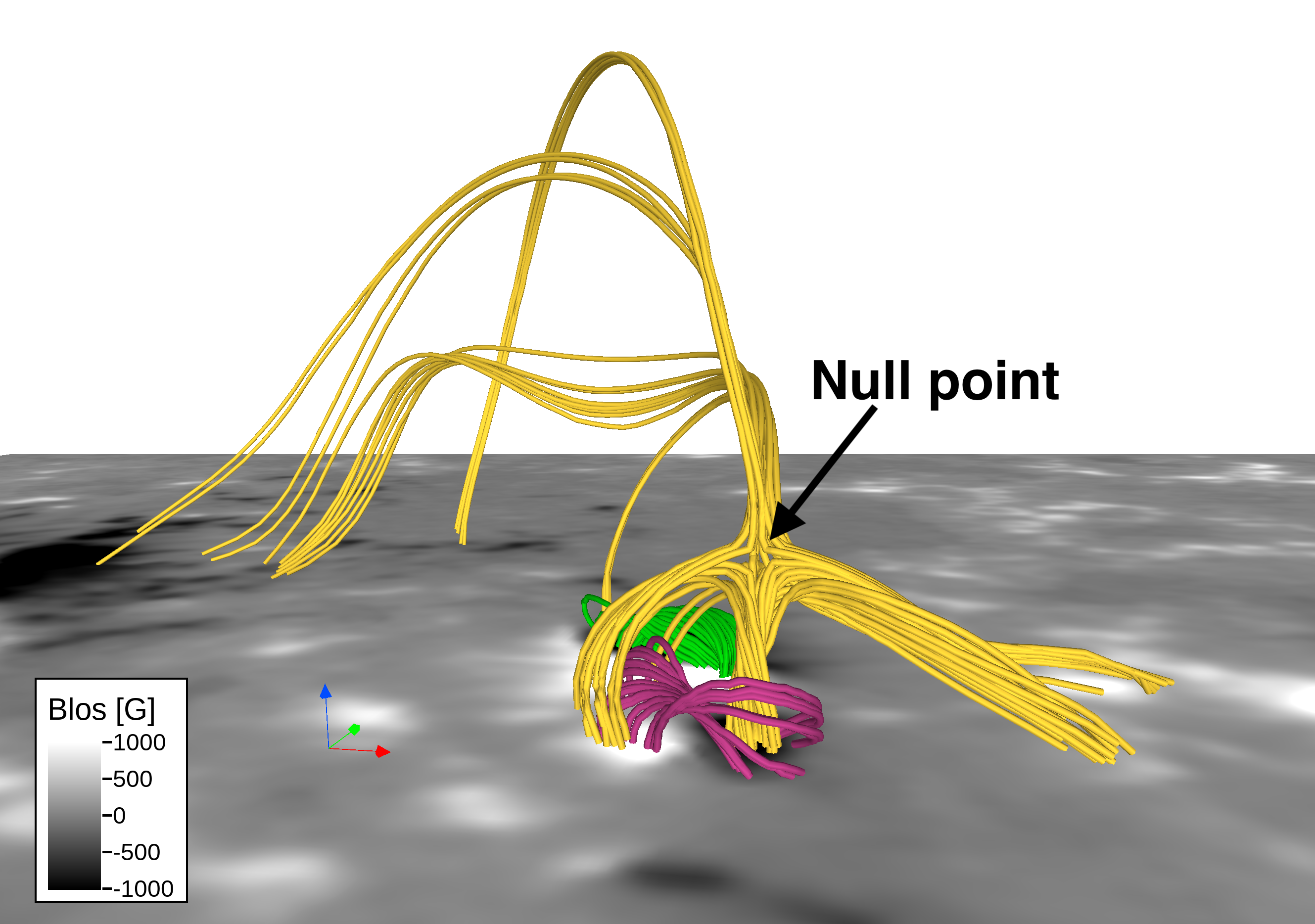}
    \includegraphics[clip,trim=0.1cm 0.45cm 0.cm 0.2cm,width=0.35\textwidth]{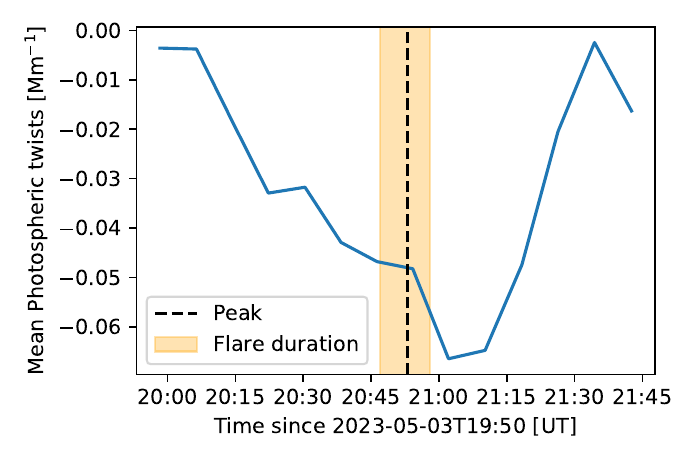}

    \caption{Magnetic field topology and mean twist evolution derived from the data-driven simulation. \textit{Left and middle panels}: top and side views of the magnetic field lines, highlighted by different colors; background image shows the HMI \Blos~at 20:54~UT. Black arrow indicates a possible location of the magnetic null point. \textit{Right panel}: the temporal evolution of the mean twist in the R1 region  of the photosphere consisting pores A and B. Yellow shaded area and dashed line refer to the GOES flare duration and flare peak time, respectively.}
    \label{fig:field_topology}
\end{figure*}

Derived from VBI images, pore B exhibits maximum preflare rotation with angular speed of approximately 34$^{\circ}$~hr$^{-1}$, which corresponds to angular acceleration of 9.2$\times10^{-8}$~rad~s$^{-2}$ within timescales of 30 minutes.
%This acceleration is higher than those observed during more intense flares (e.g., beyond C-class) in previous studies \citep{2014ApJ...782L..31W, 2016NatCo...713104L, 2016NatCo...713798B}.
What could be the cause of such a fast acceleration? Previous studies suggested that sudden photospheric motions during strong flares can be associated with changes in the Lorentz force \citep{2016NatCo...713798B, 2016NatCo...713104L}. 

To investigate this, we used Equations \ref{Equation1} and \ref{Equation2} to calculate the change in horizontal and vertical components of the Lorentz force in the photosphere, $\delta F_h$ and $\delta F_r$,  using the magnetic field vector from the ME inversion of ViSP \Fe ~data. In Figure~\ref{fig:pore_rot_flux}(d) we show the temporal evolution of $\delta F_h$ and $\delta F_r$ within the FOV encompassing both pores (R1 region highlighted in Figure~\ref{fig:pore_rot_flux}e). We find no changes in $\delta F_r$, but significant changes in $\delta F_h$ prior to the flare peak time. As demonstrated in Figure~\ref{fig:pore_rot_flux}d, we observe a 30\% relative change in  $\delta F_r$ during the flare peak. However, the limited number of observation points during the flaring period hinders confirmation of a steplike change in  $\delta F_r$. Nevertheless, a small rise during the peak suggests a potential steplike change associated with the C-class flare.% This photospheric change could be attributed to the back reaction of magnetic field line restructuring in the corona during the flare.}

To investigate whether the torque resulting from the horizontal Lorentz force is sufficient to rotate the pores, we considered both pores to have the geometry of a rigid elliptical disk rotating about its center. The torque acting on these pores from the horizontal component of the Lorentz force can be expressed as $T = I \alpha$, where $I$ and $\alpha$ represent the moment of inertia and angular acceleration, respectively. We estimate $I$ for the ellipse as $I = \frac{1}{2} \rho \pi h a b (a^2 + b^2)$, where $a$ and $b$ are the lengths of the semi-major and semi-minor axes of the ellipse, derived from ellipse shape fitting. We considered the photospheric density, $\rho$, to be $4 \times 10^{-7}$ g~cm$^{-3}$ at the photospheric height, $h$, of 270~km, as also adopted by \cite{2014ApJ...782L..31W} and \cite{2016NatCo...713104L}. 

From the observations we estimate the torque, $T = r\times \delta F_h$, where $r$ represents the distance between the minor axis edge of the ellipse and the ellipse's centroid. This results in $T = $ 3.05$\times 10^{28}$ dyne cm (mean $\approx5.3 \times 10^{28}$) for pore A and \textbf{$\approx2.95 \times 10^{28}$} dyne~cm (mean \textbf{$\approx5.2 \times 10^{28}$}) for pore B at 19:30 UT. 
These torques can produce an angular acceleration of $4.67\times10^{-6}$ (mean $\approx0.14 \times 10^{-6}$) rad s$^{-2}$ and $7.88 \times 10^{-6}$ (mean $\approx8.2 \times 10^{-6}$) rad s$^{-2}$ for pores A and B, respectively. 
These estimated values are approximately 1 order higher than the observed angular acceleration of approximately $10^{-7}$ rad s$^{-2}$. Consequently, our analysis suggests that the horizontal Lorentz force is sufficient to produce the observed angular acceleration in pores. We note, however, that the above calculations have uncertainties in the derived physical quantities due to the assumptions made regarding $h$, $\rho$, and the rigid elliptical body.

To assess overall AR activity, we used HMI SHARP data due to ViSP's limited coverage (Fig.~\ref{fig:pore_rot_flux}f, g). VBI images showed clockwise rotation in the top region (R1) and counterclockwise rotation in the pore region (R2) of the photosphere (see the online animation). Consequently, the Lorentz force derived from the HMI data, $\delta F_h$, exhibited a decreasing trend in R1 and an increasing trend in R2. 
Interestingly, the overall change in $\delta F_h$ across the entire HMI FOV (including both R1 and R2), as shown in panel (g), was dominated by the R1 region and was decreasing before the flare but showed minimal variation after the flaring period.
This suggests a potential connection between both regions and the C4.1-class flare. As suggested in previous studies, the back reaction of magnetic field line reconfiguration in the corona could be influencing the magnetic morphology in the photosphere \citep{2014ApJ...782L..31W, 2016NatCo...713104L, 2016NatCo...713798B}.

Despite a similar trend in $\delta F_h$ between ViSP and HMI in the same FOV (R1), ViSP consistently reports $\delta F_h$ values an order of magnitude lower than HMI. The cause of this discrepancy remains unclear, but several factors could potentially contribute, including differences in temporal cadence, spatial resolution, polarimetric sensitivity, varying seeing conditions of ViSP data, and the specific techniques employed by each instrument for spectropolarimetric observations. Additionally, ViSP inversions have more uncertainties in the horizontal and vertical component of the magnetic field due to varying seeing conditions.

%Additionally, we observe relatively stronger magnetic field in ViSP data compared to the HMI (see Figure~\ref{fig:visp_hmi_blos_compare}), which can further influence the measured values of $\delta F_h$, leading to the observed difference between ViSP and HMI.}

\subsection{Magnetic Field Topology Above the Active Region}
Figure~\ref{fig:field_topology} shows a 3D magnetic field structure of AR 13293 around the flare peak time ($\sim$20:54~UT) from our MF data-driven simulation (see details in Section \ref{sec:cgem_intro}). 
We find that the field lines exhibit a more sheared configuration near the PIL of pores A and B (see magenta and green field lines in Figure~\ref{fig:field_topology}) and the mixed quadrupolar region, which could be a consequence of the pores' rotation. The magnetic field configuration also includes a null point around $\sim$4~Mm above the photosphere. The fan-spine-like structure of magnetic field lines spreading out from the null point indicates that the reconnection could take place near the null point following a dome-like structure. This null point exits above two pores throughout the DKIST observations. We note that the VBI-H$\beta$ ribbon brightening is located away from this fan-like structure. The reason for the observed location of the flare ribbon away from the null point is not clear. Notably, VBI-H$\beta$ images revealed multiple small-scale brightenings near the PIL of pores A and B, as well as the sheared region, preceding the main C4.1 flare.% However, there is a possibility that most of the energy is dissipated in the corona without reaching chromospheric heights, as evidenced by the brightening above the pores in the AIA 211 Å images (see Figure~\ref{fig:hmivisp_fov}d).}

We also estimate the mean twist averaged along each field line as $\alpha=\mathrm{\vec{J}.\vec{B}/\bold{B}^2}$, where $\vec{J}$ and $\vec{B}$ are the electric current and magnetic field vectors in the simulation domain, respectively. We find that the absolute value of the mean twist shows a gradual increase before the flare, returning to its preflare value after the flare (see Figure~\ref{fig:field_topology}). The reversal in the twist rate starts roughly $\sim$10 minutes after the flare end time. Contrary to \cite{2016NatCo...713798B}, we do not find a stepwise change in the temporal profile of the mean twist. This could be attributed to a smaller flare class of our study. It is likely that the twist of the field lines arises from the rotation of the pores, powered by the changes in $\delta F_h$, which can induce instability in the field configuration. During the flaring time, the field lines may reconnect and reconfigure into a more stable state, leading to the untwisting of the field lines.

\section{Discussion}
\label{sec:Discussion}
We have analyzed the C4.1 GOES-class flare (SOL2023-05-03T20:53) using high-resolution spectropolarimetric observations recorded with the ViSP instrument and VBI at DKIST. 
High spectropolarimetric sensitivity provided a unique opportunity to investigate the relation between photospheric magnetic field and the flare. 
The magnetic structure of the AR that generated this flare is highly complex, exhibiting quadrupolar structure. The ViSP scans covered two pores of opposite polarities and a portion of another sunspot. Comparison of the \Blos~obtained from ViSP's \ion{Fe}{1} 6302~\AA\ line with SDO/HMI data showed strong agreement, with a Spearman coefficient value of 0.98. Due to its high spatial resolution and use of the full spectral profile, ViSP measures stronger \Blos~values in the pores compared to the HMI magnetogram.

High-cadence and high-resolution VBI images showed the presence of small-scale brightenings prior to the C4.1-class flare, primarily located near the PIL between two pores and in the region above (R1 and R2 regions in Figure~\ref{fig:hmivisp_fov}). 

Within the FOV of the ViSP and VBI, we observed a counterclockwise rotation of two pores. While both the pores showed rotation during the flare, pore B exhibited a significant preflare acceleration. It rotated by approximately 17$^{\circ}$ in only 30 minutes, reaching a speed of about 34$^{\circ}$~hr$^{-1}$. In contrast, pore~A showed slow increase in the rotation angle with an angular speed of $\sim$9$^{\circ}$~hr$^{-1}$. Previous observations have reported rotations of sunspot associated with X-class flare up to 50$^{\circ}$~hr$^{-1}$ \citep{2016NatCo...713104L}.
To the best of our knowledge, we report the fastest rotation of pores associated with a less intense C-class flare.

It has been speculated that the Lorentz force may be responsible for the observed rotation of sunspots in the photosphere associated with flares \citep{2008ASPC..383..221H,2009SoPh..258..203M,2014ApJ...782L..31W, 2016NatCo...713104L, 2016NatCo...713798B}. To investigate this, we estimated the Lorentz force in the photosphere using the inferred magnetic field vector from the ViSP \Fe~spectra. We observed a 30\% relative change in the horizontal component ($\delta F_h$) of Lorentz force at the flare peak time and roughly no change in the radial component ($\delta F_r$). The torque ($\sim 6.2\times 10^{28}$ dyne cm) generated from $\delta F_h$ was found to be sufficient to produce the observed angular acceleration in pores. Previous studies have reported the torque value of $\sim10^{30}$ dyne~cm associated with stronger flares \citealt{2014ApJ...782L..31W, 2016NatCo...713104L, 2016NatCo...713798B}).

As depicted in Figure~\ref{fig:pore_rot_flux}, the Lorentz force changes began after the C1.9-class flare that occurred at 18:49 UT. It is possible that the rotation of pores A and B could be related to the C1.9-class flare. However, due to limited cadence of ViSP maps, we were unable to investigate the role of this flare.

The magnetic field topology, derived from data-driven simulations, revealed a complex structure within the AR. Field lines near the PIL between pores A and B, and the region above them (R2 region in Figure~\ref{fig:pore_rot_flux}e), exhibited increased shearing and twisting. At these locations, VBI images also exhibited small-scale brightenings prior to the C4.1 flare. The data-driven simulation also revealed that the mean twist within pores in the photosphere increased before the flare and returned to its preflare value afterward. This observation contrasts with the findings of \cite{2016NatCo...713798B}, who reported a stepwise change in the temporal profile of mean photospheric twist within a rotating region during an X-class flare. They attributed this change to the shrinkage of newly formed flare loops with significant twist. 
\\
Additionally, a null point with a fan-spine structure was identified approximately 4~Mm above the photosphere. While the possibility exists that preflare pore rotation could trigger the observed flare, we were unable to find conclusive evidence to support this hypothesis. In this Letter, simulation outputs focuse solely on the analysis of the AR's magnetic topology, leaving the investigation of flare triggering and signature comparisons for future study.

Unlike M- or X-class flares typically observed by moderate telescopes with rotating sunspots and penumbrae, VBI's high resolution unveils similar rotational behavior in our C-class event, but on a significantly smaller scale. However, ViSP's 3.11-minute cadence might limit capturing rapid magnetic field variations during the flare. To fully understand flare dynamics, high spatial resolution like ViSP's needs to be combined with high temporal resolution observations achievable with DKIST's other advanced instruments.

\section{Conclusion}
\label{sec:conclusion}
High-resolution VBI images and spectropolarimetric observations from ViSP at the DKIST enabled the investigation of the relationship between rotating pores and the C4.1-class solar flare. The VBI images revealed a rapid preflare rotation of a pore at a significantly smaller scale. We found that the horizontal Lorentz force change derived from ViSP data is sufficient to explain the rotation of the observed pores. The estimated Lorentz force change in the photosphere could be caused by the back reaction of restructuring of the coronal field lines.
%, which confirms previous observation with M- or X-class flares. 
The magnetic field topology of the flaring ARs was complex and exhibiting the presence of a null-point-like configuration above rotating pores.
This analysis highlights the need for further investigation into the relationship between flare signatures observed in data-driven simulations and observations. This would help determine if the preflare pore rotation or the shearing motion of the AR plays the dominant role in triggering the flare.

\begin{acknowledgements}
We would like to thank the anonymous referee for their constructive comments and suggestions, which have significantly improved this paper. We thank the US government for providing the
funding that made this research possible. We acknowledge support
from NASA ECIP NNH18ZDA001N and NSF CAREER SPVKK1RC2MZ3.
The research reported herein is based in part on data collected with the Daniel K. Inouye Solar Telescope (DKIST) a facility of the U.S. National Science Foundation.  DKIST is operated by the National Solar Observatory under a cooperative agreement with the Association of Universities for Research in Astronomy, Inc. DKIST is located on land of spiritual and cultural significance to Native Hawaiian people. The use of this important site to further scientific knowledge is done with appreciation and respect. Funding for the DKIST Ambassadors program is provided by the National Solar Observatory, a facility of the National Science Foundation, operated under Cooperative Support Agreement number AST-1400450.
Data and images are courtesy of NASA/SDO and the HMI and AIA science teams. This research has made use of NASA’s Astrophysics Data System. We acknowledge the use of the visualization software VAPOR \citep{2019Atmos..10..488L} for generating relevant graphics. We acknowledge the community effort devoted to the development of the following open-source packages that were used in this work: NumPy (\url{numpy.org}), matplotlib (\url{matplotlib.org}) and SunPy (\url{sunpy.org}).

\end{acknowledgements}

\bibliographystyle{aa}
\bibliography{new-ref}

\begin{thebibliography}{52}
\expandafter\ifx\csname natexlab\endcsname\relax\def\natexlab#1{#1}\fi

\bibitem[{{Anwar} {et~al.}(1993){Anwar}, {Acton}, {Hudson}, {Makita}, {McClymont}, \& {Tsuneta}}]{1993SoPh..147..287A}
{Anwar}, B., {Acton}, L.~W., {Hudson}, H.~S., {et~al.} 1993, \solphys, 147, 287

\bibitem[{{Archontis} \& {T{\"o}r{\"o}k}(2008)}]{2008A&A...492L..35A}
{Archontis}, V. \& {T{\"o}r{\"o}k}, T. 2008, \aap, 492, L35

\bibitem[{{Bi} {et~al.}(2016){Bi}, {Jiang}, {Yang}, {Hong}, {Li}, {Yang}, \& {Xu}}]{2016NatCo...713798B}
{Bi}, Y., {Jiang}, Y., {Yang}, J., {et~al.} 2016, Nature Communications, 7, 13798

\bibitem[{{Carmichael}(1964)}]{1964NASSP..50..451C}
{Carmichael}, H. 1964, {A Process for Flares}, Vol.~50, 451

\bibitem[{{Cauzzi} {et~al.}(2000){Cauzzi}, {Falchi}, \& {Falciani}}]{2000A&A...357.1093C}
{Cauzzi}, G., {Falchi}, A., \& {Falciani}, R. 2000, \aap, 357, 1093

\bibitem[{{Cauzzi} {et~al.}(2008){Cauzzi}, {Reardon}, {Uitenbroek}, {Cavallini}, {Falchi}, {Falciani}, {Janssen}, {Rimmele}, {Vecchio}, \& {W{\"o}ger}}]{2008A&A...480..515C}
{Cauzzi}, G., {Reardon}, K.~P., {Uitenbroek}, H., {et~al.} 2008, \aap, 480, 515

\bibitem[{{Cheung} \& {DeRosa}(2012)}]{2012ApJ...757..147C}
{Cheung}, M. C.~M. \& {DeRosa}, M.~L. 2012, \apj, 757, 147

\bibitem[{{Chintzoglou} {et~al.}(2019){Chintzoglou}, {Zhang}, {Cheung}, \& {Kazachenko}}]{2019ApJ...871...67C}
{Chintzoglou}, G., {Zhang}, J., {Cheung}, M. C.~M., \& {Kazachenko}, M. 2019, \apj, 871, 67

\bibitem[{{da Silva Santos} {et~al.}(2023){da Silva Santos}, {Reardon}, {Cauzzi}, {Schad}, {Mart{\'\i}nez Pillet}, {Tritschler}, {W{\"o}ger}, {Hofmann}, {Stauffer}, \& {Uitenbroek}}]{2023ApJ...954L..35D}
{da Silva Santos}, J.~M., {Reardon}, K., {Cauzzi}, G., {et~al.} 2023, \apjl, 954, L35

\bibitem[{{de la Cruz Rodr{\'\i}guez} \& {van Noort}(2017)}]{2017SSRv..210..109D}
{de la Cruz Rodr{\'\i}guez}, J. \& {van Noort}, M. 2017, \ssr, 210, 109

\bibitem[{{de Wijn} {et~al.}(2022){de Wijn}, {Casini}, {Carlile}, {Lecinski}, {Sewell}, {Zmarzly}, {Eigenbrot}, {Beck}, {W{\"o}ger}, \& {Kn{\"o}lker}}]{2022SoPh..297...22D}
{de Wijn}, A.~G., {Casini}, R., {Carlile}, A., {et~al.} 2022, \solphys, 297, 22

\bibitem[{{Fisher} {et~al.}(2012){Fisher}, {Bercik}, {Welsch}, \& {Hudson}}]{2012SoPh..277...59F}
{Fisher}, G.~H., {Bercik}, D.~J., {Welsch}, B.~T., \& {Hudson}, H.~S. 2012, \solphys, 277, 59

\bibitem[{{Fisher} {et~al.}(2020){Fisher}, {Kazachenko}, {Welsch}, {Sun}, {Lumme}, {Bercik}, {DeRosa}, \& {Cheung}}]{2020ApJS..248....2F}
{Fisher}, G.~H., {Kazachenko}, M.~D., {Welsch}, B.~T., {et~al.} 2020, \apjs, 248, 2

\bibitem[{{Fletcher} {et~al.}(2011){Fletcher}, {Dennis}, {Hudson}, {Krucker}, {Phillips}, {Veronig}, {Battaglia}, {Bone}, {Caspi}, {Chen}, {Gallagher}, {Grigis}, {Ji}, {Liu}, {Milligan}, \& {Temmer}}]{2011SSRv..159...19F}
{Fletcher}, L., {Dennis}, B.~R., {Hudson}, H.~S., {et~al.} 2011, \ssr, 159, 19

\bibitem[{{Gary} \& {Hagyard}(1990)}]{1990SoPh..126...21G}
{Gary}, G.~A. \& {Hagyard}, M.~J. 1990, \solphys, 126, 21

\bibitem[{{Hirayama}(1974)}]{1974SoPh...34..323H}
{Hirayama}, T. 1974, \solphys, 34, 323

\bibitem[{{Hoeksema} {et~al.}(2020){Hoeksema}, {Abbett}, {Bercik}, {Cheung}, {DeRosa}, {Fisher}, {Hayashi}, {Kazachenko}, {Liu}, {Lumme}, {Lynch}, {Sun}, \& {Welsch}}]{2020ApJS..250...28H}
{Hoeksema}, J.~T., {Abbett}, W.~P., {Bercik}, D.~J., {et~al.} 2020, \apjs, 250, 28

\bibitem[{{Hudson} {et~al.}(2008){Hudson}, {Fisher}, \& {Welsch}}]{2008ASPC..383..221H}
{Hudson}, H.~S., {Fisher}, G.~H., \& {Welsch}, B.~T. 2008, in Astronomical Society of the Pacific Conference Series, Vol. 383, Subsurface and Atmospheric Influences on Solar Activity, ed. R.~{Howe}, R.~W. {Komm}, K.~S. {Balasubramaniam}, \& G.~J.~D. {Petrie}, 221

\bibitem[{{Kazachenko} {et~al.}(2022){Kazachenko}, {Albelo-Corchado}, {Tamburri}, \& {Welsch}}]{2022SoPh..297...59K}
{Kazachenko}, M.~D., {Albelo-Corchado}, M.~F., {Tamburri}, C.~A., \& {Welsch}, B.~T. 2022, \solphys, 297, 59

\bibitem[{{Kazachenko} {et~al.}(2014){Kazachenko}, {Fisher}, \& {Welsch}}]{2014ApJ...795...17K}
{Kazachenko}, M.~D., {Fisher}, G.~H., \& {Welsch}, B.~T. 2014, \apj, 795, 17

\bibitem[{{Kopp} \& {Pneuman}(1976)}]{1976SoPh...50...85K}
{Kopp}, R.~A. \& {Pneuman}, G.~W. 1976, \solphys, 50, 85

\bibitem[{{Kuhn} {et~al.}(1994){Kuhn}, {Balasubramaniam}, {Kopp}, {Penn}, {Dombard}, \& {Lin}}]{1994SoPh..153..143K}
{Kuhn}, J.~R., {Balasubramaniam}, K.~S., {Kopp}, G., {et~al.} 1994, \solphys, 153, 143

\bibitem[{{Kuridze} {et~al.}(2016){Kuridze}, {Mathioudakis}, {Christian}, {Kowalski}, {Jess}, {Grant}, {Kawate}, {Sim{\~o}es}, {Allred}, \& {Keenan}}]{2016ApJ...832..147K}
{Kuridze}, D., {Mathioudakis}, M., {Christian}, D.~J., {et~al.} 2016, \apj, 832, 147

\bibitem[{{Leenaarts} {et~al.}(2010){Leenaarts}, {Rutten}, {Reardon}, {Carlsson}, \& {Hansteen}}]{2010ApJ...709.1362L}
{Leenaarts}, J., {Rutten}, R.~J., {Reardon}, K., {Carlsson}, M., \& {Hansteen}, V. 2010, \apj, 709, 1362

\bibitem[{{Leka} {et~al.}(2014){Leka}, {Barnes}, \& {Crouch}}]{2014ascl.soft04007L}
{Leka}, K.~D., {Barnes}, G., \& {Crouch}, A. 2014, {AMBIG: Automated Ambiguity-Resolution Code}, Astrophysics Source Code Library, record ascl:1404.007

\bibitem[{{Lemen} {et~al.}(2012){Lemen}, {Title}, {Akin}, {Boerner}, {Chou}, {Drake}, {Duncan}, {Edwards}, {Friedlaender}, {Heyman}, {Hurlburt}, {Katz}, {Kushner}, {Levay}, {Lindgren}, {Mathur}, {McFeaters}, {Mitchell}, {Rehse}, {Schrijver}, {Springer}, {Stern}, {Tarbell}, {Wuelser}, {Wolfson}, {Yanari}, {Bookbinder}, {Cheimets}, {Caldwell}, {Deluca}, {Gates}, {Golub}, {Park}, {Podgorski}, {Bush}, {Scherrer}, {Gummin}, {Smith}, {Auker}, {Jerram}, {Pool}, {Soufli}, {Windt}, {Beardsley}, {Clapp}, {Lang}, \& {Waltham}}]{2012SoPh..275...17L}
{Lemen}, J.~R., {Title}, A.~M., {Akin}, D.~J., {et~al.} 2012, \solphys, 275, 17

\bibitem[{{Li} {et~al.}(2019){Li}, {Jaroszynski}, {Pearse}, {Orf}, \& {Clyne}}]{2019Atmos..10..488L}
{Li}, S., {Jaroszynski}, S., {Pearse}, S., {Orf}, L., \& {Clyne}, J. 2019, Atmosphere, 10, 488

\bibitem[{{Liu} {et~al.}(2016){Liu}, {Xu}, {Cao}, {Deng}, {Lee}, {Hudson}, {Gary}, {Wang}, {Jing}, \& {Wang}}]{2016NatCo...713104L}
{Liu}, C., {Xu}, Y., {Cao}, W., {et~al.} 2016, Nature Communications, 7, 13104

\bibitem[{{Louis} {et~al.}(2015){Louis}, {Kliem}, {Ravindra}, \& {Chintzoglou}}]{2015SoPh..290.3641L}
{Louis}, R.~E., {Kliem}, B., {Ravindra}, B., \& {Chintzoglou}, G. 2015, \solphys, 290, 3641

\bibitem[{{Lumme} {et~al.}(2022){Lumme}, {Pomoell}, {Price}, {Kilpua}, {Kazachenko}, {Fisher}, \& {Welsch}}]{Lumme2022}
{Lumme}, E., {Pomoell}, J., {Price}, D.~J., {et~al.} 2022, \aap, 658, A200

\bibitem[{{Metcalf}(1994)}]{1994SoPh..155..235M}
{Metcalf}, T.~R. 1994, \solphys, 155, 235

\bibitem[{{Min} \& {Chae}(2009)}]{2009SoPh..258..203M}
{Min}, S. \& {Chae}, J. 2009, \solphys, 258, 203

\bibitem[{{Moore} {et~al.}(2001){Moore}, {Sterling}, {Hudson}, \& {Lemen}}]{2001ApJ...552..833M}
{Moore}, R.~L., {Sterling}, A.~C., {Hudson}, H.~S., \& {Lemen}, J.~R. 2001, \apj, 552, 833

\bibitem[{{Pesnell} {et~al.}(2012){Pesnell}, {Thompson}, \& {Chamberlin}}]{2012SoPh..275....3P}
{Pesnell}, W.~D., {Thompson}, B.~J., \& {Chamberlin}, P.~C. 2012, \solphys, 275, 3

\bibitem[{{Petrie}(2019)}]{2019ApJS..240...11P}
{Petrie}, G. J.~D. 2019, \apjs, 240, 11

\bibitem[{{Qiu} {et~al.}(2023){Qiu}, {Alaoui}, {Antiochos}, {Dahlin}, {Swisdak}, {Drake}, {Robison}, {DeVore}, \& {Uritsky}}]{2023ApJ...955...34Q}
{Qiu}, J., {Alaoui}, M., {Antiochos}, S.~K., {et~al.} 2023, \apj, 955, 34

\bibitem[{{Rempel} {et~al.}(2023){Rempel}, {Chintzoglou}, {Cheung}, {Fan}, \& {Kleint}}]{2023ApJ...955..105R}
{Rempel}, M., {Chintzoglou}, G., {Cheung}, M. C.~M., {Fan}, Y., \& {Kleint}, L. 2023, \apj, 955, 105

\bibitem[{{Rimmele} {et~al.}(2020){Rimmele}, {Warner}, {Keil}, {Goode}, {Kn{\"o}lker}, {Kuhn}, {Rosner}, {McMullin}, {Casini}, {Lin}, {W{\"o}ger}, {von der L{\"u}he}, {Tritschler}, {Davey}, {de Wijn}, {Elmore}, {Fehlmann}, {Harrington}, {Jaeggli}, {Rast}, {Schad}, {Schmidt}, {Mathioudakis}, {Mickey}, {Anan}, {Beck}, {Marshall}, {Jeffers}, {Oschmann}, {Beard}, {Berst}, {Cowan}, {Craig}, {Cross}, {Cummings}, {Donnelly}, {de Vanssay}, {Eigenbrot}, {Ferayorni}, {Foster}, {Galapon}, {Gedrites}, {Gonzales}, {Goodrich}, {Gregory}, {Guzman}, {Guzzo}, {Hegwer}, {Hubbard}, {Hubbard}, {Johansson}, {Johnson}, {Liang}, {Liang}, {McQuillen}, {Mayer}, {Newman}, {Onodera}, {Phelps}, {Puentes}, {Richards}, {Rimmele}, {Sekulic}, {Shimko}, {Simison}, {Smith}, {Starman}, {Sueoka}, {Summers}, {Szabo}, {Szabo}, {Wampler}, {Williams}, \& {White}}]{2020SoPh..295..172R}
{Rimmele}, T.~R., {Warner}, M., {Keil}, S.~L., {et~al.} 2020, \solphys, 295, 172

\bibitem[{{Sainz Dalda}(2017)}]{2017ApJ...851..111S}
{Sainz Dalda}, A. 2017, \apj, 851, 111

\bibitem[{{Sanchez Almeida} \& {Lites}(1992)}]{1992ApJ...398..359S}
{Sanchez Almeida}, J. \& {Lites}, B.~W. 1992, \apj, 398, 359

\bibitem[{{Scherrer} {et~al.}(2012){Scherrer}, {Schou}, {Bush}, {Kosovichev}, {Bogart}, {Hoeksema}, {Liu}, {Duvall}, {Zhao}, {Title}, {Schrijver}, {Tarbell}, \& {Tomczyk}}]{2012SoPh..275..207S}
{Scherrer}, P.~H., {Schou}, J., {Bush}, R.~I., {et~al.} 2012, \solphys, 275, 207

\bibitem[{{Schrijver}(2009)}]{2009AdSpR..43..739S}
{Schrijver}, C.~J. 2009, Advances in Space Research, 43, 739

\bibitem[{{Sturrock}(1966)}]{1966Natur.211..695S}
{Sturrock}, P.~A. 1966, \nat, 211, 695

\bibitem[{{Sudol} \& {Harvey}(2005)}]{2005ApJ...635..647S}
{Sudol}, J.~J. \& {Harvey}, J.~W. 2005, \apj, 635, 647

\bibitem[{{Sun} {et~al.}(2017){Sun}, {Hoeksema}, {Liu}, {Kazachenko}, \& {Chen}}]{2017ApJ...839...67S}
{Sun}, X., {Hoeksema}, J.~T., {Liu}, Y., {Kazachenko}, M., \& {Chen}, R. 2017, \apj, 839, 67

\bibitem[{{Toriumi} \& {Wang}(2019)}]{2019LRSP...16....3T}
{Toriumi}, S. \& {Wang}, H. 2019, Living Reviews in Solar Physics, 16, 3

\bibitem[{{T{\"o}r{\"o}k} {et~al.}(2024){T{\"o}r{\"o}k}, {Linton}, {Leake}, {Miki{\'c}}, {Lionello}, {Titov}, \& {Downs}}]{2024ApJ...962..149T}
{T{\"o}r{\"o}k}, T., {Linton}, M.~G., {Leake}, J.~E., {et~al.} 2024, \apj, 962, 149

\bibitem[{{Wang} {et~al.}(2014){Wang}, {Liu}, {Deng}, \& {Wang}}]{2014ApJ...782L..31W}
{Wang}, S., {Liu}, C., {Deng}, N., \& {Wang}, H. 2014, \apjl, 782, L31

\bibitem[{{W{\"o}ger} {et~al.}(2021){W{\"o}ger}, {Rimmele}, {Ferayorni}, {Beard}, {Gregory}, {Sekulic}, \& {Hegwer}}]{2021SoPh..296..145W}
{W{\"o}ger}, F., {Rimmele}, T., {Ferayorni}, A., {et~al.} 2021, \solphys, 296, 145

\bibitem[{{Yadav} {et~al.}(2021){Yadav}, {D{\'\i}az Baso}, {de la Cruz Rodr{\'\i}guez}, {Calvo}, \& {Morosin}}]{2021A&A...649A.106Y}
{Yadav}, R., {D{\'\i}az Baso}, C.~J., {de la Cruz Rodr{\'\i}guez}, J., {Calvo}, F., \& {Morosin}, R. 2021, \aap, 649, A106

\bibitem[{{Yadav} \& {Kazachenko}(2023)}]{2023ApJ...944..215Y}
{Yadav}, R. \& {Kazachenko}, M.~D. 2023, \apj, 944, 215

\bibitem[{{Yadav} {et~al.}(2017){Yadav}, {Mathew}, \& {Tiwary}}]{2017SoPh..292..105Y}
{Yadav}, R., {Mathew}, S.~K., \& {Tiwary}, A.~R. 2017, \solphys, 292, 105

\end{thebibliography}

\end{document}